\documentclass[conf]{new-aiaa}
\usepackage[utf8]{inputenc}

\usepackage{graphicx}
\usepackage{amsmath}
\usepackage[version=4]{mhchem}
\usepackage{siunitx}
\usepackage{longtable,tabularx}
\usepackage{subfigure}
\usepackage{multirow}

\title{Atmospheric Density Uncertainty Quantification for Satellite Conjunction Assessment}

\author{David J. Gondelach\footnote{Postdoctoral Associate, Department of Aeronautics and Astronautics, 77 Massachusetts Avenue, Cambridge, MA. E-mail: dgondela@mit.edu} and Richard Linares\footnote{Charles Stark Draper Assistent Professor, Department of Aeronautics and Astronautics, 77 Massachusetts Avenue, Cambridge, MA, Senior Member AIAA. E-mail: linaresr@mit.edu}}
\affil{Massachusetts Institute of Technology, Cambridge, MA 02139}

\begin{document}

\maketitle

\begin{abstract}
Conjunction assessment requires knowledge of the uncertainty in the predicted orbit. Errors in the atmospheric density are a major source of error in the prediction of low Earth orbits. Therefore, accurate estimation of the density and quantification of the uncertainty in the density is required. Most atmospheric density models, however, do not provide an estimate of the uncertainty in the density. In this work, we present a new approach to quantify uncertainties in the density and to include these for calculating the probability of collision $P_c$. For this, we employ a recently developed dynamic reduced-order density model that enables efficient prediction of the thermospheric density. First, the model is used to obtain accurate estimates of the density and of the uncertainty in the estimates. Second, the density uncertainties are propagated forward simultaneously with orbit propagation to include the density uncertainties for $P_c$ calculation. For this, we account for the effect of cross-correlation in position uncertainties due to density errors on the $P_c$. Finally, the effect of density uncertainties and cross-correlation on the $P_c$ is assessed.
The presented approach provides the distinctive capability to quantify the uncertainty in atmospheric density and to include this uncertainty for conjunction assessment while taking into account the dependence of the density errors on location and time. In addition, the results show that it is important to consider the effect of cross-correlation on the $P_c$, because ignoring this effect can result in severe underestimation of the collision probability.
\end{abstract}

\section{Introduction}
The growing population of objects in orbit around Earth, due to increasing number of satellites and orbital debris, increases the spatial density and therefore the probability of collision between objects. Collision with an active satellite can result in reduced operational performance or even complete destruction of the spacecraft, thereby generating new space debris. Avoiding collision between two objects requires knowledge of their orbits. Therefore, satellites and other objects need to be track and their orbits predicted to detect close encounters. For conjunctions that have the potential to result in collision (often a miss distance less than 10 km is used as threshold), the risk of collision needs to be assessed. 

Accurate calculating of the probability of collision $P_c$ requires both accurate prediction of the orbit and accurate estimation of the uncertainty in the orbit prediction. The main sources of error in orbit prediction are errors in observational data, limited knowledge of object properties (shape, mass and material) and attitude, and errors in the modelling of the atmosphere needed for drag calculations (National Research Council, 2012). 

Uncertainties in the initial state due to measurement errors can be estimated during orbit determination. In addition, uncertainties in the drag coefficient, frontal area and mass can be combined in the ballistic coefficient, which can be solved for in orbit determination and can be included in the covariance. 
On the other hand, errors in the atmospheric density estimates are often unknown or not well characterized. Vallado and Finkleman \cite{vallado2014critical} reported an average one sigma accuracy of 10\% to 15\% for empirical density models. However, the error in density depends on the employed density model as well as on the solar activity and location \cite{he2018review}. Therefore, there is a strong need to both accurately estimate density and the uncertainty in the density. The goal of this work is to quantify the uncertainty in atmospheric density and to take these uncertainties into account for collision probability calculations. 

The current state of practice by the U.S. Strategic Command for incorporating uncertainty in the atmospheric density is via the use of a consider parameter. In this approach, the covariance does not contain the atmospheric density uncertainty directly; instead, the estimated uncertainty in the density is added to the ballistic coefficient variance. 
Because atmospheric density and the ballistic coefficient are multiplicatively coupled, the ballistic coefficient variance, which is part of the covariance, can be changed to achieve an effect equivalent to considering the density uncertainty separately. Here, the percent error in atmospheric density forecast is obtained by comparing the predictions from the Jacchia-Bowman-HASDM-2009 (JBH09) model with the true densities computed using corrections from the High Accuracy Satellite Drag Model (HASDM) \cite{poore2016covariance}. One drawback of this approach is that by incorporating the density error in the ballistic coefficient variance one scales density values equally in all atmospheric regions, whereas in reality the density error is not uniform over the density field. This aspect becomes important when the conjunction between two significantly different orbits (e.g. different mean altitude or different orbital plane) are considered.

Several studies have looked into the effect of uncertainty in the atmospheric density in the orbit prediction \cite{anderson2009sensitivity,leonard2012impact,emmert2014propagation,emmert2017propagation,sagnieres2017uncertainty,SCHIEMENZ2019grid,SCHIEMENZ2019Least,SCHIEMENZ2019Prop}; however, few have considered the effect on the probability of collision. In addition, most of these studies focused on the error in atmospheric density due to inaccurate space weather inputs and forecasts, because these errors become dominant for orbit prediction of several days. However, for orbit determination and short-term orbit predictions, the error in the density model itself is expected to be larger. Therefore, density models need to be calibrated and the uncertainties in the model must be estimated. Several techniques exist for density model calibration \cite{cefola2004atmospheric,yurasov2005density,storz2005high,doornbos2008use,chen2019improved}. However, these techniques do not provide direct estimates of the uncertainty in the density. In addition, the calibration is carried out using static atmospheric models and, therefore, it is not possible to predict how uncertainties in the density evolve over time (only the statistics of historical predictions can be studied). There is, therefore, a strong need for the quantification of uncertainty in density models and the propagation of these uncertainties into the future.

Two studies that considered the effect of density errors on the collision probability were carried by Bussy-Virat et al.\cite{bussy2018effects} and Hejduk and Snow \cite{Hejduk2018Effect}.
Bussy-Virat et al.\cite{bussy2018effects} analysed the effect of errors in the future space weather prediction on the $P_c$ and calculated the $P_c$ for different future space weather scenarios to obtain a probability density function (PDF) for the $P_c$. For this, they considered errors in the solar activity proxies $F10.7$ and $A_p$, but did not consider errors in the density model itself. Hejduk and Snow \cite{Hejduk2018Effect} studied how the severity level of conjunctions changes due to errors in density and due to including density uncertainties for $P_c$ calculations. They assumed density errors of a factor 0.5 to 2.0 with respect to the model density due to model and space weather forecast errors and used a consider parameter to include density uncertainties in the covariance. They analysed the effect of the error and uncertainties on the $P_c$ and showed that ignoring density uncertainties can lead to failure to detect severe conjunction events. Hejduk and Snow did however not consider the cross-correlation in position uncertainties due to density errors.

If an estimate of the uncertainty in density is available, one can study the effect of density errors on the probability of collision. For this, different technique for computing the $P_c$ can be employed. One approach is to consider a Monte Carlo process in which ensemble orbits are computed based on sampling from probability density functions of the assumed uncertainties. The $P_c$ is then given by the number of collisions between the two ensembles divided by the total number of possible scenarios. The approach is computationally expensive because it requires the computation of thousands of ensemble orbits in order to obtain an accurate $P_c$ estimate.
Other widely-used techniques, called linear or 2D $P_c$ estimation techniques, can be used when the duration of the encounter is short, such that the position uncertainties can be assumed to be constant during the close approach. This methodology involves computing the covariance matrices at the time of closest approach. The covariances of the two objects, which represent a Gaussian approximation of the probability density function of the states, can be combined to calculate the covariance of the miss distance. The miss distance uncertainty (i.e. the combined covariance matrix) is then compared to the combined hard-body radius circle placed at relative position vector at TCA to compute the probability of collision.

It is important to note that if the joint position covariance is computed for calculating the probability of collision, one must account for correlation in the position errors \cite{coppola2004effects,casali2018effect}. In other words, if covariances contain a correlated error, then one needs to remove this error in order to compute the correct combined covariance. Otherwise, the combined covariance is too large, which affects the estimated $P_c$. If two objects fly through the same atmosphere, the position errors due to errors in the density will be correlated. Therefore, accounting for the cross-correlation is important (see \cite{coppola2004effects,casali2018effect}); however, most previous work does not consider this effect on the $P_c$ calculation. 

In this paper, we propose a new technique to estimate and include the uncertainty in atmospheric density for CA. For this, we use a recently developed dynamic density model that enables density forecasting \cite{mehta2018quasi}. Common physics-based models of the thermosphere, such as GITM and TIE-GCM, require solving Navier-Stokes equations over a high-dimensional discretized spatial grid involving $10^4$-$10^6$ state variables, which is computationally expensive. Mehta et al. \cite{mehta2018quasi} used reduced-order modelling to develop a low-dimensional dynamic model for the thermospheric density. This model enables efficient estimation of the density using a Kalman filter. Density estimation was demonstrated using accelerometer-derived density measurements \cite{mehta2018new}, using GPS-derived orbit data \cite{mehta2018data} and using two-line element data \cite{gondelach2019real}. The authors showed that density estimates based two-line element data were more accurate than the JB2008 and NRLMSISE-00 models \cite{gondelach2019real}. In addition, the estimates for the uncertainty in estimated density were obtained. These density uncertainties depend on the latitude, local solar time and altitude. In addition, the uncertainty was found to be dependent on the epoch (i.e. solar activity) and measurement data. This is a significant improvement over a single global density error estimate that does not include location and time dependence of density errors. Moreover, the density prediction depends on the solar activity, such that also the predicted density uncertainty depends on the future space weather. 

We use the dynamic model to estimate both the global atmospheric density using two-line element data and the uncertainty in the estimated density. The dynamic model is then used to predict the future density and corresponding uncertainty. By propagating an orbit and the atmospheric state simultaneously and by employing uncertainty propagation, we obtain the uncertainty in the orbit due to uncertainty in both the initial orbital state and in the density model. Finally, we use the orbit and covariance computed for two objects to calculate the probability of collision between the objects and compare the result with the $P_c$ in case density uncertainties are not considered. Here, the effect of cross-correlation due to density errors is considered as well.

In previous work, the authors demonstrated the use of the dynamic density model for accurate density estimation and density uncertainty quantification \cite{gondelach2019real}. The contributions of this work are:
\begin{enumerate}
\item The future density uncertainty is computed by propagating the uncertainty during the density prediction.
\item The orbital state uncertainty that includes uncertainty due to the density errors is computed by propagating the state and density uncertainties simultaneously.
\item The probability of collision is computed using position covariances that includes uncertainty due to density.
\item The combined position covariance used for collision probability calculation is adjusted to account for the effect of cross-correlation in the position uncertainties due to density errors.
\item The effect of density uncertainties and the effect of cross-correlation on the probability of collision estimate is assessed.
\item The estimated collision probability is compared with collision probabilities computing using Monte Carlo analysis.
\end{enumerate}

In the following, first the methods employed for the new approach are discussed. After that the test cases for conjunction assessment are presented. The $P_c$ is computed for the different scenarios with and without considering density uncertainties and cross-correlation effects. Finally, the results are discussed and conclusions are drawn.

\section{Methodology}
\label{Methodology}
First, we briefly introduce the development of the dynamic reduced-order density model. After that, density estimation using the dynamic model and two-line elements data with a Kalman filter is discussed. Finally, the technique of computing the collision probability in the work is presented.

\subsection{Dynamic reduced-order density model}
The goal of the development of a dynamic reduced-order density model is to enable computationally-efficient forecasting of the atmospheric density. This is achieved by reducing the dimension of atmospheric state with respect to physics-based models and by deriving a dynamic model for the reduced-order state.

The main idea of reduced-order modeling is to reduce the dimensionality of the state space while retaining maximum information. In our case, the full state space consists of the neutral mass density values on a dense uniform grid in latitude, local solar time and altitude. 
First, to make the problem tractable, the state space dimension is reduced using Proper Orthogonal Decomposition (POD). Second, a linear dynamic model is derived by applying Dynamic Mode Decomposition with control (DMDc).

\subsubsection{Proper orthogonal decomposition} 
The concept of order reduction using POD is to project the high-dimensional system and its solution onto a set of low-dimensional basis functions or spatial modes, while capturing the dominant characteristics of the system. 
Consider the variation $\tilde{\mathbf{x}}$ of the neutral mass density $\mathbf{x}$ with respect to the mean value $\bar{\mathbf{x}}$:
\begin{equation}
    \tilde{\mathbf{x}}(\mathbf{s},t) = {\mathbf{x}}(\mathbf{s},t) - \bar{\mathbf{x}}(\mathbf{s})
    \label{eq:variationX}
\end{equation}
where $\mathbf{s}$ is the spatial grid. Here, we use density in the log scale because its variance is more uniform, such that:
\begin{equation}
    \mathbf{x} = \log_{10}\boldsymbol{\rho},~~~~ \bar{\mathbf{x}} = \log_{10}\bar{\boldsymbol{\rho}},~~~~ \tilde{\mathbf{x}} = \log_{10}{{\boldsymbol{\rho}}} - \log_{10}\bar{\boldsymbol{\rho}}
\end{equation}
A significant fraction of the variance $\tilde{\mathbf{x}}$ can be captured by the first $r$ principal spatial modes:
\begin{equation}
    \tilde{\mathbf{x}}(\mathbf{s},t) \approx \sum_{i=1}^r c_i(t) \Phi_i(\mathbf{s})
\end{equation}
where $\Phi_i$ are the spatial modes and $c_i$ are the corresponding time-dependent coefficients.
The spatial modes $\Phi$ are computed using a singular value decomposition (SVD) of the snapshot matrix ${\bf X}$ that contains $\tilde{\mathbf{x}}$ for different times:
\begin{equation}
{\bf X}= \begin{bmatrix}
| & | & & | \\
\tilde{\mathbf{x}}_1 & \tilde{\mathbf{x}}_2 & \cdots & \tilde{\mathbf{x}}_m \\
| & | & & |
\end{bmatrix}
= \mathbf{U} \mathbf{\Sigma} \mathbf{V}^T
\label{eq:SVD}
\end{equation}
The spatial or POD modes $\Phi$ are given by the left singular vectors (the columns of $\mathbf{U}$).

The state reduction is achieved using a similarity transform:
\begin{equation}
    {\mathbf{z}} = \mathbf{U}_r^{-1} \tilde{\mathbf{x}}  = \mathbf{U}_r^T \tilde{\mathbf{x}} 
    \label{eq:orderReduction}
\end{equation}
where $\mathbf{U}_r$ is a matrix with the first $r$ POD modes and ${\mathbf{z}}$ is our reduced-order state. Projecting ${\mathbf{z}}$ back to the full space gives approximately $\tilde{\mathbf{x}}$ that allows us to compute the density:
\begin{equation}
    {\mathbf{x}}(\mathbf{s},t) \approx \mathbf{U}_r(\mathbf{s})\,\mathbf{z}(t) + \bar{\mathbf{x}}(\mathbf{s})
    \label{eq:densityFromROM}
\end{equation}
More details on POD can be found in \cite{mehta2017methodology}.

\subsubsection{Dynamic Mode Decomposition with control}
To enable prediction of the atmospheric density, we develop a linear dynamic model for the reduced-order state $\mathbf{z}$:
\begin{equation}
    \mathbf{z}_{k+1} = \mathbf{A} \mathbf{z}_k+ \mathbf{B} \mathbf{u}_k
    \label{eq:linearModel}
\end{equation}
where $\mathbf{u}_k$ is the system input, which in our case are the space weather inputs. The dynamic matrix ${\bf A}$ and input matrix $\mathbf{B}$ can be estimated from output data (i.e. historical or simulated density and space weather data) using the DMDc algorithm. Details of the DMDc technique for deriving the dynamic density model can be found in \cite{mehta2018quasi}.

To enable continuous-time propagation of the ROM modes for estimation, the discrete-time matrices are converted to continuous-time matrices (see \cite{mehta2018quasi}) such that we obtain the reduced-order density dynamics:
\begin{equation}
    \dot{\mathbf{z}} = \mathbf{A}_c \mathbf{z}+ \mathbf{B}_c \mathbf{u}
    \label{eq:contlinearModel}
\end{equation}
where ${\bf A}_c$ is the dynamic matrix and ${\bf B}_c$ is the input matrix in continuous time. This equation is used to propagate the density state for estimation, prediction and uncertainty propagation.

\subsubsection{JB2008-based ROM density model}
In this work, we use a reduced-order density model based on density data from the JB2008 density model. We computed hourly density data on a dense grid in latitude, local solar time and altitude over a period of 12 years from 1999 to 2010. The data was then used to perform POD and DMDc to obtain our dynamic reduced-order model for the thermosphere. Details of the model can be found in \cite{gondelach2019real}.

\subsection{Density estimation}
Once we have obtained our dynamic reduced-order density model, Eq.~\eqref{eq:contlinearModel}, we can estimate the the reduced-order density state $z$ using a Kalman filter. In this work, this is achieved by simultaneously estimating the reduced-order density state $z$ and the orbital states of objects by data assimilation of two-line element orbital data as described in a previous paper by the authors \cite{gondelach2019real}. The estimation process has the following characteristics:
\begin{itemize}
\item \textbf{Filter}: We use a square-root unscented Kalman filter (UKF) for estimation \cite{wan2001unscented}
\item \textbf{State}: The state $\textbf{x}$ that is estimated consists of the osculating orbital states (expressed in modified equinoctial elements (MEE)), the BCs of the objects and the reduced-order density state ${\bf z}$:
\begin{equation}
\textbf{x}= \begin{bmatrix}
p_1, f_1, g_1, h_1, k_1, L_1,{BC_1}, & ... &, p_n, f_n, g_n, h_n, k_n, L_n,{BC_n}, & {\bf z}^T \\
\end{bmatrix} ^T
\end{equation}
where $n$ is the number of objects.
\item \textbf{Measurements}: The measurements consist of osculating orbital states extracted from TLE data (expressed in MEE). 
\item \textbf{Dynamical model}: The reduced-order density state $z$ is propagated using Eq.~\eqref{eq:contlinearModel}. The orbits are propagated in the J2000 reference frame considering geopotential acceleration (EGM2008 model, 20x20 harmonics) and atmospheric drag computed using the ROM density model.
\end{itemize}
For more details about the orbital dynamical model and process and measurement noise used for estimation we refer the reader to \cite{gondelach2019real}.

\subsection{Uncertainty estimation}
From the Kalman filter estimation, we obtain both an estimate for the state \textbf{x} and an estimate of the covariance $\mathbf{P}_{\textbf{x}}$:
\begin{equation*}
\mathbf{P}_{\textbf{x}} = 
\begin{bmatrix}
    \sigma_{p_1}^2 & \sigma_{p_1f_1} & \sigma_{p_1g_1} & \ldots & \ldots & \ldots & \ldots \\
    \sigma_{p_1f_1} & \sigma_{f_1}^2 & \sigma_{f_1g_1} & \ldots & \ldots & \ldots & \ldots \\
    \sigma_{p_1g_1} & \sigma_{f_1g_1} & \sigma_{g_1}^2 & \ldots & \ldots & \ldots & \ldots \\
    \vdots & \vdots & \vdots & \ddots & \ldots & \ldots & \ldots \\
    \vdots & \vdots & \vdots & \vdots & \sigma_{z_1}^2 & \ldots& \sigma_{z_1z_r} \\
    \vdots & \vdots & \vdots & \vdots & \vdots & \ddots& \vdots \\
    \vdots & \vdots & \vdots & \vdots & \sigma_{z_1z_r} & \ldots & \sigma_{z_r}^2
\end{bmatrix}
\end{equation*}
The lower right part of $\mathbf{P}_{\textbf{x}}$ contains the covariance of the reduced-order density state $\textbf{z}$, $\mathbf{P}_{\textbf{z}}$. This covariance can be used to compute the effect of density uncertainty on the position uncertainty and thus on the collision probability. In addition, $\mathbf{P}_{\textbf{z}}$ can be converted to uncertainty in the physical density. Using a single row of $\mathbf{U}_r$ that corresponds to one grid point $i$ in the full space, $\mathbf{U}_{r,i}$, we obtain the uncertainty in the density at the grid point $\rho_i$ as follows:

\begin{align}
    \sigma_{\tilde{x}_i}^2 &= \mathbf{U}_{r,i}~ \mathbf{P}_{\textbf{z}} ~\mathbf{U}_{r,i}^T \label{eq:xtildeuncertainty} \\
    \sigma_{\rho_i} &= \rho_i ~\ln{(10)} ~\sigma_{\tilde{x}_i} \label{eq:rhouncertainty}
\end{align}
where $\sigma_{\tilde{x}_i}^2$ is the variance of the variation of density $\tilde{x}$ at grid point $i$. The percent uncertainty is then calculated as:
\begin{equation}
    \sigma_{\rho_i}^{\%} = \frac{\sigma_{\rho_i}}{\rho_i} \cdot 100 \% =  ~\ln{(10)} ~\sigma_{\tilde{x}_i}  \cdot 100 \% \label{eq:rhouncertaintyPerc}
\end{equation}

\subsection{Collision probability calculation}
We use Alfano's method \cite{alfano2005numerical} to calculate the probability of collision $P_c$. This technique provides an accurate estimate of the true $P_c$ if the PDF of the position is Gaussian and duration of the encounter between the two object is very short such that the PDF is constant during the encounter. In this work, we have assessed the Gaussianity and short encounter duration for our test cases. In addition, we will assume that the distance of close approach (DCA) is not affected by the error in the density, whereas in reality it is. However, if the change in DCA due to uncertainties is small compared to the covariance size, then the effect on the $P_c$ is small.

Alfano's method requires the following inputs for each of the two objects (with $i=1,2$ for the primary and secondary object, respectively):
\begin{itemize}
\item Position vector at TCA: $\mathbf{r}_i=[r_{x_i},r_{y_i},r_{z_i}]$
\item Velocity vector at TCA: $\mathbf{v}_i=[v_{x_i},v_{y_i},v_{z_i}]$
\item Position covariance matrix at TCA: $\mathbf{P}_i$
\item Object body radius $R_i$
\end{itemize}

The $P_c$ is then computed by projecting the combined position covariance and relative miss distance on the so-called encounter plane and computing the probability that the miss distance is smaller than the combined body radius:
\begin{itemize}
\item \textbf{Step 1}: Calculate the encounter plane, i.e. the plane perpendicular to the relative velocity $\mathbf{v}_r = \mathbf{v}_2 - \mathbf{v}_1$.
\item \textbf{Step 2}: Project the combined covariance $\mathbf{P}_{m} = \mathbf{P}_1 + \mathbf{P}_2$ and relative distance $\mathbf{r}_m = \mathbf{r}_2 - \mathbf{r}_1$ on the encounter plane.
\item \textbf{Step 3}: Calculate $P_c$ by computing the integral of the two-dimensional PDF in the encounter plane over the circular region defined by the cross-section of the combined object.
\end{itemize}
The integral for computing the probability of collision is given by \cite{Hemenway2014Achieving}:
\begin{equation}
    P_c = \frac{1}{\sqrt{8\pi}\sigma_x} \int^R_{0} \left[ \mathrm{erf} \left( \frac{y_m + \sqrt{R^2-x^2}}{\sqrt{2} \sigma_y} \right) + \left( \frac{-y_m + \sqrt{R^2-x^2}}{\sqrt{2} \sigma_y} \right)\right] \left[ \exp{\left( \frac{-(x+x_m)^2}{2\sigma_x^2} \right)} + \exp{\left( \frac{-(-x+x_m)^2}{2\sigma_x^2} \right)} \right] dx
\end{equation}
where 
\begin{equation}
    \mathrm{erf}(z) = \frac{2}{\sqrt{\pi}} \int^z_0 e^{-t^2} dt
\end{equation}
is the error function and $R=R_1+R_2$ is the combined body radius of the two objects. For the integral, the $x$-axis is aligned with the major axis of the projected covariance such that the projected covariance is given by the standard deviations $\sigma_x$ and $\sigma_y$ and the relative position is given by $x_m$ and $y_m$. This integral can be computed numerically to obtain $P_c$.
More information on Alfano's method can be found in \cite{alfano2005numerical} and \cite{Hemenway2014Achieving}. 

In the Section~\ref{Methodology}\ref{sec:crosscorr}, we will show how to compute the combined covariance $\mathbf{P}_{m}$ that accounts for cross-correlation between $\mathbf{P}_1$ and $\mathbf{P}_2$ due to density errors.

\subsection{Uncertainty propagation}
To compute the $P_c$, we need to calculate the position covariances $\mathbf{P}_1$ and $\mathbf{P}_2$ at the TCA. For this, we propagate the covariance matrix $\mathbf{P}_{\textbf{x}}$ using an unscented transformation (i.e. sigma point propagation) that is also used in the UKF \cite{julier1997new,wan2001unscented}. Since, the covariance $\mathbf{P}_{\textbf{x}}$ includes the orbital state, BC and reduced-order state $z$, the effect of uncertainties in each of these are included in the position covariance at TCA. 

For uncertainty propagation, the orbital state is expressed in modified equinoctial elements (MEE) to mitigate the departure from ``Gaussianity''  of the state PDF under the non-linear propagation with respect to Cartesian coordinates. After computing the covariance in MEE at TCA, the covariance is converted to Cartesian space using unscented transformation in order to obtain the position covariance needed in Alfano's method. Note that if the position PDF becomes strongly non-Gaussian, then Alfano's method does not provide an accurate approximation of the $P_c$ anymore.

\subsection{Cross-correlation}
\label{sec:crosscorr}
Casali et al. \cite{casali2018effect} described a technique to consider the effect of cross-correlation of orbital errors on the probability of collision. In this section, we follow the derivation by Casali et al. \cite{casali2018effect} to obtain a equation to correct for cross-correlation due to density errors in the calculation of $P_c$.

Starting from the definition of the miss distance $\mathbf{r}_m = \mathbf{r}_2 - \mathbf{r}_1$, the mean and covariance of the miss distance are given by:
\begin{equation}
    \boldsymbol{\mu}_m = \mathrm{E}[\mathbf{r}_m] = \mathrm{E}[\mathbf{r}_2 - \mathbf{r}_1] = \mathrm{E}[\mathbf{r}_2] - \mathrm{E}[\mathbf{r}_1] = \boldsymbol{\mu}_2 - \boldsymbol{\mu}_1
\end{equation}
\begin{align}
    \mathbf{P}_m = & \mathrm{E}[(\mathbf{r}_m-\boldsymbol{\mu}_m)(\mathbf{r}_m-\boldsymbol{\mu}_m)^T] \\
    = & \mathrm{E}[((\mathbf{r}_2-\boldsymbol{\mu}_2)-(\mathbf{r}_1-\boldsymbol{\mu}_1))((\mathbf{r}_2-\boldsymbol{\mu}_2)-(\mathbf{r}_1-\boldsymbol{\mu}_1))^T] \\
    = & \mathrm{E}[(\mathbf{r}_2-\boldsymbol{\mu}_2)(\mathbf{r}_2-\boldsymbol{\mu}_2)^T] + \mathrm{E}[(\mathbf{r}_1-\boldsymbol{\mu}_1)(\mathbf{r}_1-\boldsymbol{\mu}_1)^T] - \mathrm{E}[(\mathbf{r}_2-\boldsymbol{\mu}_2)(\mathbf{r}_1-\boldsymbol{\mu}_1)^T] - \mathrm{E}[(\mathbf{r}_1-\boldsymbol{\mu}_1)(\mathbf{r}_2-\boldsymbol{\mu}_2)^T] \label{eq:pmCross} \\
    = & \mathbf{P}_2 + \mathbf{P}_1 - \mathrm{E}[(\mathbf{r}_2-\boldsymbol{\mu}_2)(\mathbf{r}_1-\boldsymbol{\mu}_1)^T] - \mathrm{E}[(\mathbf{r}_1-\boldsymbol{\mu}_1)(\mathbf{r}_2-\boldsymbol{\mu}_2)^T]
\end{align}
When the position errors in $\mathbf{r}_1$ and $\mathbf{r}_2$ are independent, the cross-correlation terms are zero, such that $\mathbf{P}_m = \mathbf{P}_2 - \mathbf{P}_1$. This is generally the assumption when performing conjunction assessment using linear techniques, because errors from orbit determination are assumed to statistically independent. 

Now, let us assume that the errors in the orbit predictions are due to both errors in the initial state $\delta$ and global model parameter errors $\mathbf{\Delta}_g$. Then, for the position deviations at TCA we can write:
\begin{align}
    \mathbf{r}_1 - \boldsymbol{\mu}_1 &= \mathbf{\Phi}_1 \boldsymbol{\delta}_1 + \mathbf{G}_1 \mathbf{\Delta}_g \\
    \mathbf{r}_2 - \boldsymbol{\mu}_2 &= \mathbf{\Phi}_2 \boldsymbol{\delta}_2 + \mathbf{G}_2 \mathbf{\Delta}_g
\end{align}
where $\mathbf{\Phi}$ is the state transition matrix that maps the deviation in the initial state $\boldsymbol{\delta}$ to a deviation in the state at TCA, and $\mathbf{G}$ is a state transition matrix that provides the linear approximation of the effect of the error $\mathbf{\Delta}_g$ on the state at TCA over the prediction window, like $\mathbf{\Phi}$ does. Using these expressions for the state deviations, we obtain for Eq.~\eqref{eq:pmCross}:
\begin{align}
    \mathbf{P}_m &= \mathrm{E}[(\mathbf{\Phi}_2 \boldsymbol{\delta}_2 + \mathbf{G}_2 \mathbf{\Delta}_g)(\mathbf{\Phi}_2 \boldsymbol{\delta}_2 + \mathbf{G}_2 \mathbf{\Delta}_g)^T] + \mathrm{E}[(\mathbf{\Phi}_1 \boldsymbol{\delta}_1 + \mathbf{G}_1 \mathbf{\Delta}_g)(\mathbf{\Phi}_1 \boldsymbol{\delta}_1 + \mathbf{G}_1 \mathbf{\Delta}_g)^T] \\
    &- \mathrm{E}[(\mathbf{\Phi}_2 \boldsymbol{\delta}_2 + \mathbf{G}_2 \mathbf{\Delta}_g)(\mathbf{\Phi}_1 \boldsymbol{\delta}_1 + \mathbf{G}_1 \mathbf{\Delta}_g)^T] - \mathrm{E}[(\mathbf{\Phi}_1 \boldsymbol{\delta}_1 + \mathbf{G}_1 \mathbf{\Delta}_g)(\mathbf{\Phi}_2 \boldsymbol{\delta}_2 + \mathbf{G}_2 \mathbf{\Delta}_g)^T] \label{eq:pmCross2}
\end{align}
Now we assume there is no dependence between the errors due OD and between the error due to OD and the global error (i.e. $\mathrm{E}[\boldsymbol{\delta}_1\boldsymbol{\delta}_2^T]=0$ and $\mathrm{E}[\boldsymbol{\delta}_1\mathbf{\Delta}_g^T]=\mathrm{E}[\boldsymbol{\delta}_2\mathbf{\Delta}_g^T]=0$), whereas $\mathrm{E}[\mathbf{\Delta}_g \mathbf{\Delta}_g^T] = \mathbf{P}_g$. We then get:
\begin{align}
    \mathbf{P}_m &= \mathbf{\Phi}_2 \mathbf{P}_{2,0} \mathbf{\Phi}_2^T + \mathbf{\Phi}_1 \mathbf{P}_{1,0} \mathbf{\Phi}_1^T + (\mathbf{G}_2 \mathbf{P}_g \mathbf{G}_2^T + \mathbf{G}_1 \mathbf{P}_g \mathbf{G}_1^T - \mathbf{G}_2 \mathbf{P}_g \mathbf{G}_1^T - \mathbf{G}_1 \mathbf{P}_g \mathbf{G}_2^T) \\
    &= (\mathbf{\Phi}_2 \mathbf{P}_{2,0} \mathbf{\Phi}_2^T + \mathbf{G}_2 \mathbf{P}_g \mathbf{G}_2^T) + (\mathbf{\Phi}_1 \mathbf{P}_{1,0} \mathbf{\Phi}_1^T + \mathbf{G}_1 \mathbf{P}_g \mathbf{G}_1^T) - \mathbf{G}_2 \mathbf{P}_g \mathbf{G}_1^T - \mathbf{G}_1 \mathbf{P}_g \mathbf{G}_2^T \\
    &= \mathbf{P}_2 + \mathbf{P}_1 - \mathbf{G}_2 \mathbf{P}_g \mathbf{G}_1^T - \mathbf{G}_1 \mathbf{P}_g \mathbf{G}_2^T \label{eq:combinedPcrosscorr}
\end{align}
where $\mathbf{P}_{2,0}$ and $\mathbf{P}_{1,0}$ are the initial covariances representing errors in $\boldsymbol{\delta}$.

In our case, the global errors $\mathbf{\Delta}_g$ are given by the covariance of the reduced-order density state, $\mathbf{P}_{\mathbf{z}}$. Therefore, to calculate the cross-correlation in the state errors due $\mathbf{\Delta}_g$, we need to compute the matrices $\mathbf{G}_1$ and $\mathbf{G}_2$ that give the first-order relation between errors in the density and in the states at TCA. 

The matrix $\mathbf{G}$ is a Jacobian that contains the partial derivatives of the position $\mathbf{r}$ at TCA with respect to the reduced-order density state at the initial epoch $\mathbf{z}_0$:
\begin{equation}
\mathbf{G} = \frac{\partial \mathbf{r}}{\partial \mathbf{z}_0} = 
\begin{bmatrix}
    \frac{\partial {r}_x}{\partial {z}_{0,1}} & \ldots & \frac{\partial {r}_x}{\partial {z}_{0,r}} \\
    \vdots & \ddots & \vdots \\
    \frac{\partial {r}_z}{\partial {z}_{0,1}} & \ldots & \frac{\partial {r}_z}{\partial {z}_{0,r}}
\end{bmatrix}
\end{equation}
This partial derivatives are approximated using central finite differences. For this, we use the sigma points that we already computed for propagating the covariance, such that the partials are approximated as, e.g.:
\begin{equation}
\frac{\partial {r}_x}{\partial {z}_{0,1}} \approx \frac{{r}_x(\Delta {z}_{0,1}) - {r}_x(-\Delta {z}_{0,1})}{2 \Delta {z}_{0,1}}
\end{equation}
where ${r}_x(\Delta {z}_{0,1})$ is the value of ${r}_x$ when a change of $\Delta {z}_{0,1}$ is applied to the initial state. These values are taking from the results of the sigma point propagation.

Once $\mathbf{G}_1$ and $\mathbf{G}_2$ are computed, we can calculate the covariance of the miss distance $\mathbf{P}_m$ that accounts for cross-correlation due to atmospheric density errors using Eq.~\eqref{eq:combinedPcrosscorr}. This $\mathbf{P}_m$ is then used in Alfano's method to calculate the $P_c$ that accounts for the effect of cross-correlation.

\section{Test cases}
To test our new approach for computing the effect of density uncertainties on $P_c$, we generate artificial conjunction scenarios. 
The test case consists of two nearly-identical orbits at about 400 km altitude that have a different longitude of the node, such that they have a conjunction point at high latitude, see Table~\ref{tab:testcase}. 
The conjunction is set to take place on February 13, 2003 at 0:00:00 UTC. The conjunction assessment is carried out over a two-day window such that the start epoch for orbit prediction for $P_c$ estimation is two days earlier on February 11, 2003 at 0:00:00 UTC. 

The atmospheric density state at the start epoch was obtained from density estimation between February 1 and 11 using TLE data, as described in \cite{gondelach2019real}. The reduced-order density state was then propagated forward for two days to obtain a prediction of the atmospheric density, which is the nominal density for the conjunction assessment.

The initial conditions of the orbits two days before conjunction are obtained by starting from the conjunction point and propagating the orbits backward in time to the start epoch (i.e. two days before the time of conjunction) using the nominal atmospheric density. 
Six different conjunction scenarios were created by varying the BC of one of the objects and by changing the DCA (by changing the in-track position of one of the objects at TCA), see Table~\ref{tab:testcase2}. The combined hard-body radius $R$ was set to 2 m.

\begin{table}
\centering
\begin{tabular}{lcccccc}
\hline
Orbit & $a$ [km] & $e$ [-] & $i$ [deg] & $\Omega$ [deg] & $\omega$ [deg] & $\nu$ [deg] \\
\hline
1 & 6778.13630 & 0.00300 & 89.0 & 0.0 & 90.0 & 0.41418532 \\
2 & 6778.13630 & 0.00300 & 89.0 & 45.0 & 90.0 & -0.41418532 \\
\hline
\end{tabular}
\caption{Nominal orbits at collision point.}
\label{tab:testcase}
\end{table}

\begin{table}
\centering
\begin{tabular}{lclcc}
\hline
Scenario & Test case ID & DCA [km] & BC$_1$ [m$^2$/kg] & BC$_2$ [m$^2$/kg] \\
\hline
\multirow{3}{*}{Same BC} & S0 & 0.00031 & 0.01 & 0.01 \\
 & S1 & 0.92375 & 0.01 & 0.01 \\
 & S2 & 1.84756 & 0.01 & 0.01 \\
\cline{1-5}
\multirow{3}{*}{Different BC} & D0 & 0.00015 & 0.01 & 0.001 \\
 & D1 & 0.92412 & 0.01 & 0.001 \\
 & D2 & 1.84756 & 0.01 & 0.001 \\
\hline
\end{tabular}
\caption{Six conjunction scenarios with different distance of close approach (DCA) and object ballistic coefficient (BC).}
\label{tab:testcase2}
\end{table}

For the conjunction assessment, we assume uncertainties in the initial states, BCs and reduced-order density.
The assumed uncertainty in the initial states expressed in MEE is:
\begin{equation}
\begin{bmatrix}
\sigma_p \\ \sigma_f \\ \sigma_g \\ \sigma_h \\ \sigma_k \\ \sigma_L
\end{bmatrix}
= 
\begin{bmatrix}
0.000140546843775429 \\
1.54877138747695E-07 \\
1.54788273902825E-07 \\
2.50908224756075E-07 \\
1.36880945273941E-07 \\
2.28039102756189E-06
\end{bmatrix}
\end{equation}
These uncertainties were obtained from OD for the GRACE-A satellite using TLE data and subsequently scaled by factor 0.1 (i.e. reduced by one order of magnitude) to obtain magnitudes for the uncertainty that are more realistic when having access to accurate orbital data.
The BC is assumed to have a 1-$\sigma$ error of 0.5\%: $\sigma_{BC} = 0.005\cdot BC$.

The initial uncertainty in the density is the uncertainty in the estimated density on February 11, 2003 at midnight that was obtained using the TLE data of 17 objects from February 1 to February 11, 2003.
The estimated uncertainty in the density at 400 and 500 km is shown in Figure~\ref{fig:densityCov} in the Results section.

The $P_c$ is computed using Alfano's method with and without considering density uncertainties and with and without considering cross-correlation for computing the combined position covariance. Finally, to estimate the true $P_c$ in case of density uncertainty, we perform Monte Carlo analyses by sampling initial densities from the density PDF and calculating the resulting $P_c$ considering only initial state uncertainty. From the Monte Carlo analyses we obtain a mean $P_c$ that is the expected $P_c$ value.

\section{Results}
\label{sec:results}

\subsection{Density estimation and uncertainty quantification}
\begin{figure}
     \centering
     \includegraphics[width=\textwidth,trim={0cm 0cm 0cm 0cm},clip]{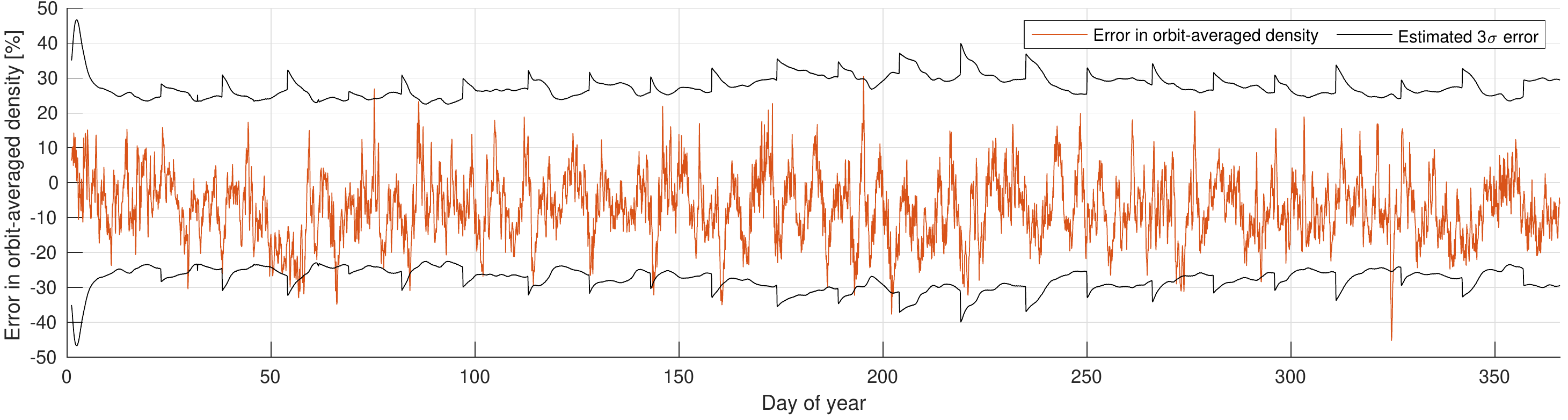}
     \caption{Error in ROM-estimated orbit-averaged density and estimated 3-$\sigma$ error along CHAMP's orbit in 2007.}
     \label{fig:2003estimated3sigma}
\end{figure}

The reduced-order density model based on JB2008 density data was used in previous work by the authors to estimate the density in the years 2003 and 2007 \cite{gondelach2019real}. Table~\ref{tab:densityErrorsStatistics} shows the statistics of the errors in the NRLMSISE-00, JB2008 and ROM-estimated densities with respect to CHAMP and GRACE-A accelerometer-derived densities in 2003 and 2007. The results show that the estimation approach performs well in debiasing the densities and the standard deviation of the error in estimated density is significantly lower than for the empirical model densities. Therefore, improved estimates for the thermospheric density can be obtained using the reduced-order density model and two-line element data. In addition, Figure~\ref{fig:2003estimated3sigma} shows the error and estimated 3-$\sigma$ error in orbit-averaged density along CHAMP's orbit over the year 2007. (The estimated 3-$\sigma$ error jumps up every 15 days, because the density estimation was restarted every 15 days to enable parallel calculation. After a couple of days of estimation, the estimated density becomes more accurate and estimated uncertainty drops.) 98.6\% of the errors are within the estimated 3-$\sigma$ error bounds. This show that estimated uncertainty provides a good estimate of the true error in density.

\begin{table}
\centering
\begin{tabular}{llcccc}
\hline
Satellite  & Model & \multicolumn{4}{c}{Density error [\%]} \\ 
  &  & \multicolumn{2}{c}{2003}  & \multicolumn{2}{c}{2007} \\ \cline{3-6}
 &  & $\mu$ & $\sigma$ & $\mu$ & $\sigma$  \\
 \hline
\multirow{3}{*}{CHAMP} & NRLMSISE-00 & 4.4 & 19.7 & 26.4 & 15.4 \\
 & JB2008 & -1.9 & 12.5 & 9.5 & 13.4 \\
 & ROM & -5.6 & 11.3 & -7.3 & 9.4 \\
 \hline
\multirow{3}{*}{GRACE-A} & NRLMSISE-00 & 20.7 & 28.6 & 44.1 & 28.4 \\
 & JB2008 & 10.0 & 18.3 & 16.0 & 28.4 \\
 & ROM & 5.9 & 16.9 & -2.6 & 22.2 \\
\hline
\end{tabular}
\caption{Mean $\mu$ and standard deviation $\sigma$ of orbit-averaged density error [\%]}
\label{tab:densityErrorsStatistics}
\end{table}

\begin{figure}
     \centering
     \subfigure[][Uncertainty in estimated density]{\includegraphics[width=0.40\textwidth,trim={0 0 16cm 0},clip]{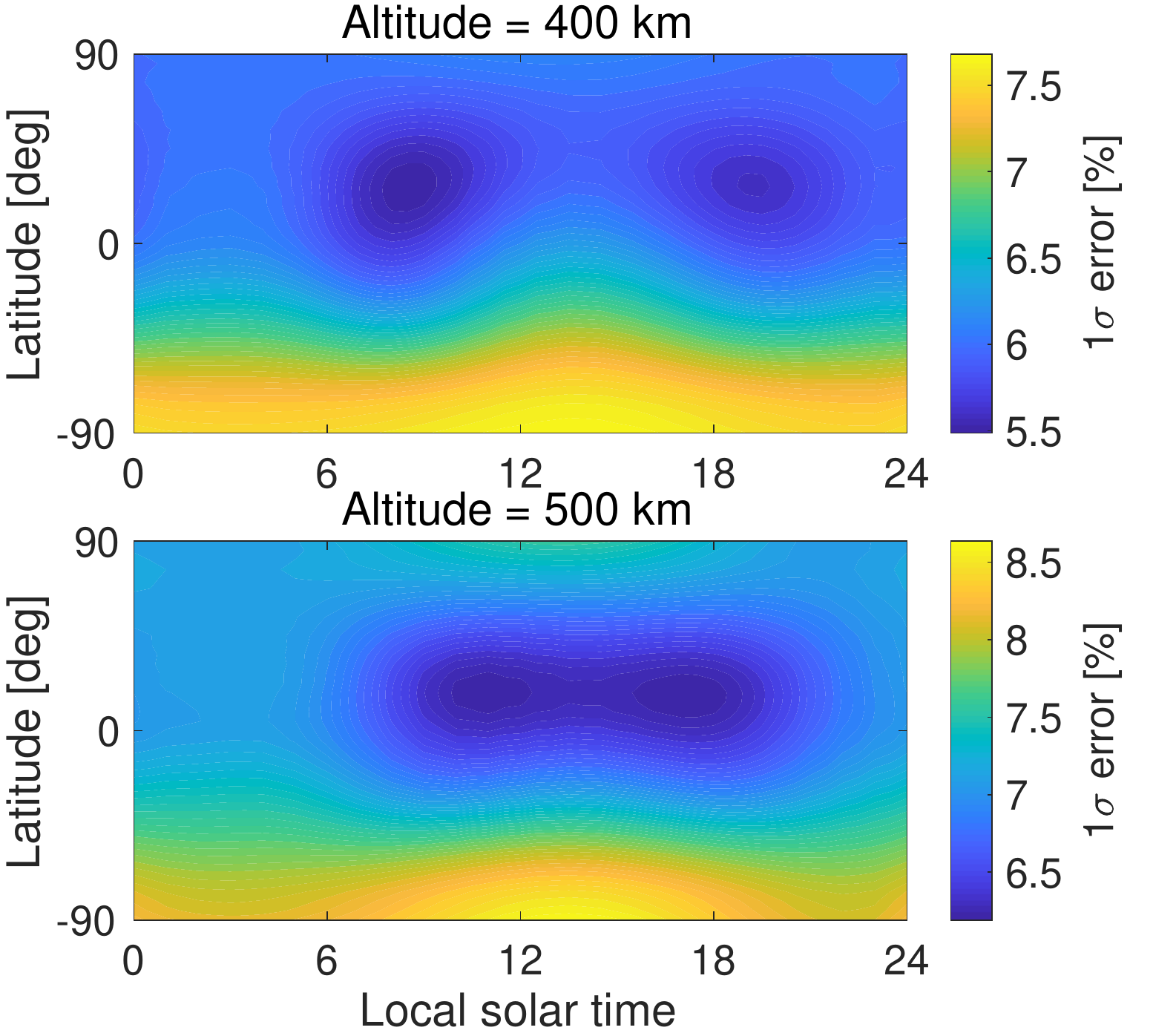} \label{fig:densityCov_est}}
     \hspace{20pt}%
     \subfigure[][Uncertainty in 2-day predicted density]{\includegraphics[width=0.40\textwidth,trim={16cm 0 0 0},clip]{DensityErrorLatLstMaps_ROM_190911034707_after10dayEstimation_after2dayPrediction.pdf} \label{fig:densityCov_pred}}
     
     \caption{1-$\sigma$ uncertainty in density at 400 and 500 km altitude: a) uncertainty in estimated density after 10 days estimation on February 11, 2003; b) uncertainty in predicted density after 2 days prediction on February 13, 2003.}
     \label{fig:densityCov}
\end{figure}

The uncertainty in the estimated density on February 11, 2003 at 400 and 500 km altitude is shown in Figure~\ref{fig:densityCov_est}. Here, the uncertainties were computed using Eqs.~\eqref{eq:xtildeuncertainty} and \eqref{eq:rhouncertaintyPerc}. 
For the conjunction assessment, the density and uncertainty were propagated for two days up to the time of closest approach. In this 2-day window, the solar activity was moderately high; the F10.7 was about 135 and the Ap varied between 5 and 27. Figure~\ref{fig:densityCov_pred} shows the uncertainty after 2-day prediction. One can see that the uncertainties have grown as a result of the atmospheric dynamics and external forcing by solar activity. 

\begin{figure}
     \centering
     
     \subfigure[][Density]{\includegraphics[width=\textwidth]{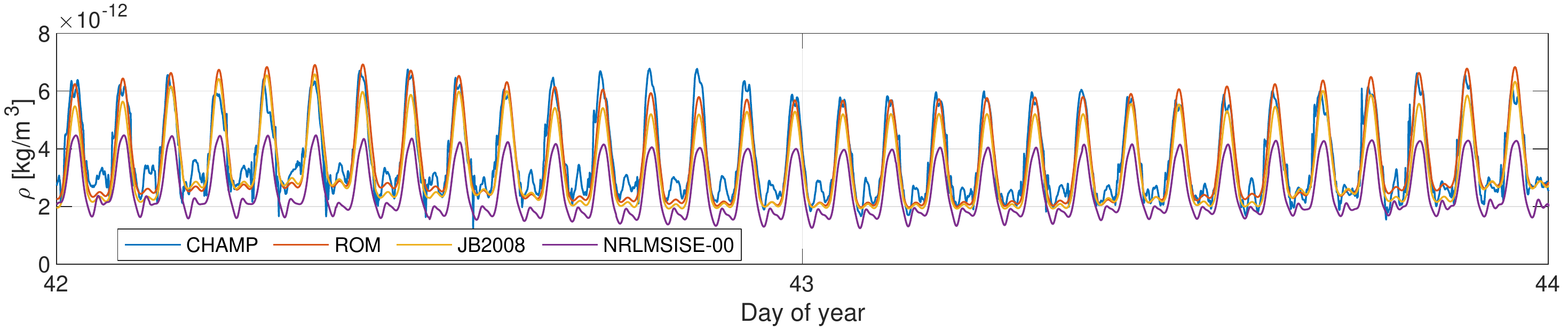} \label{fig:champDensityPred_all}}

     \subfigure[][$3\sigma$ uncertainty in ROM-predicted density]{\includegraphics[width=\textwidth]{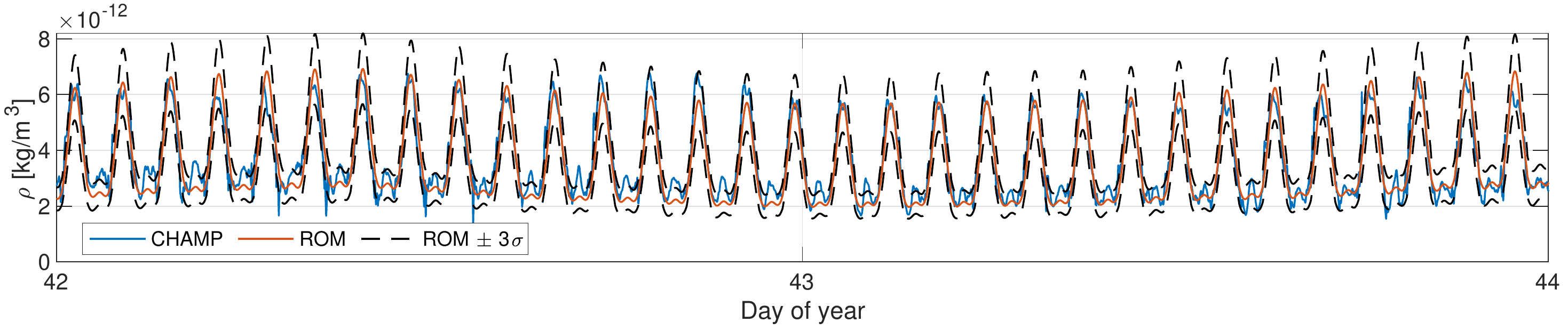} \label{fig:champDensityPred_cov}}
     
     \subfigure[][Orbit-averaged density]{\includegraphics[width=0.48\textwidth]{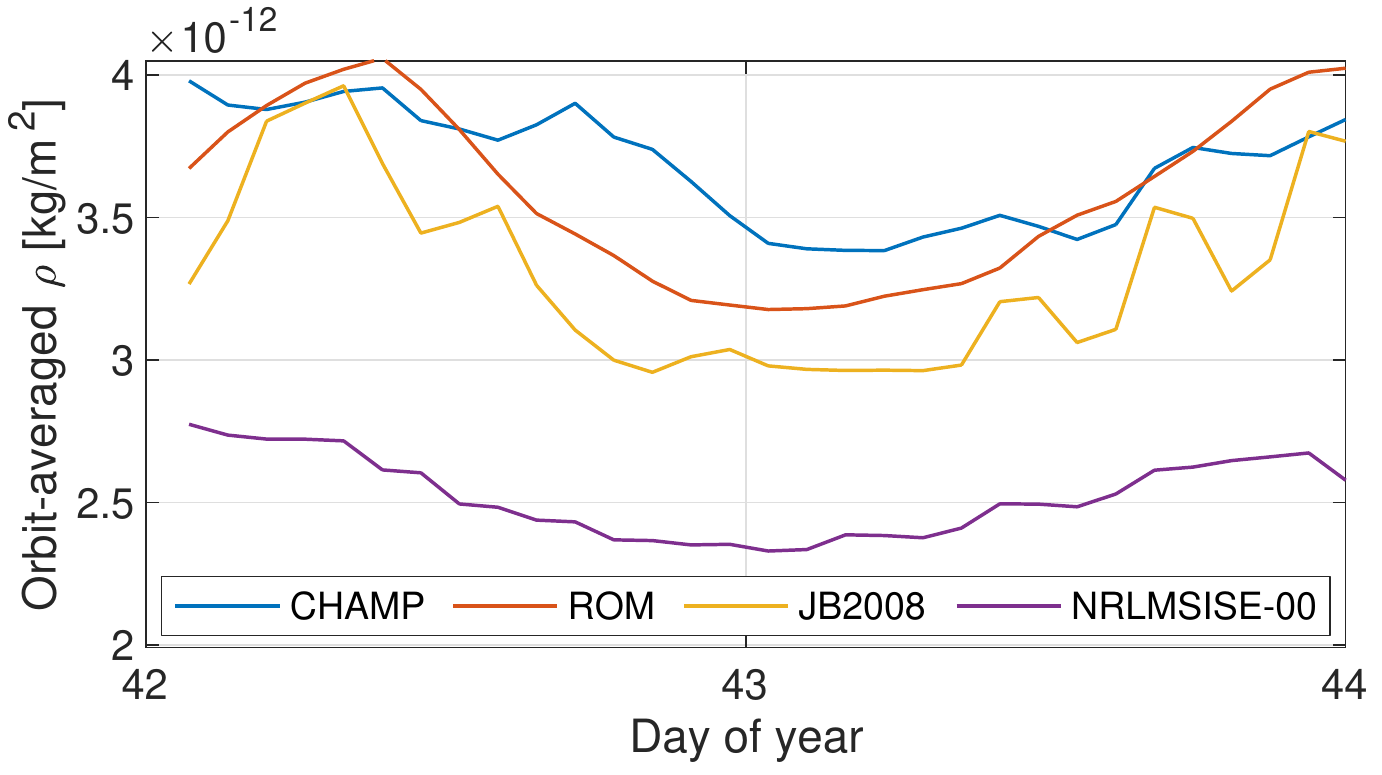} \label{fig:champDensityPred_avg}}
     \hspace{5pt}%
     \subfigure[][Orbit-averaged density error]{\includegraphics[width=0.48\textwidth]{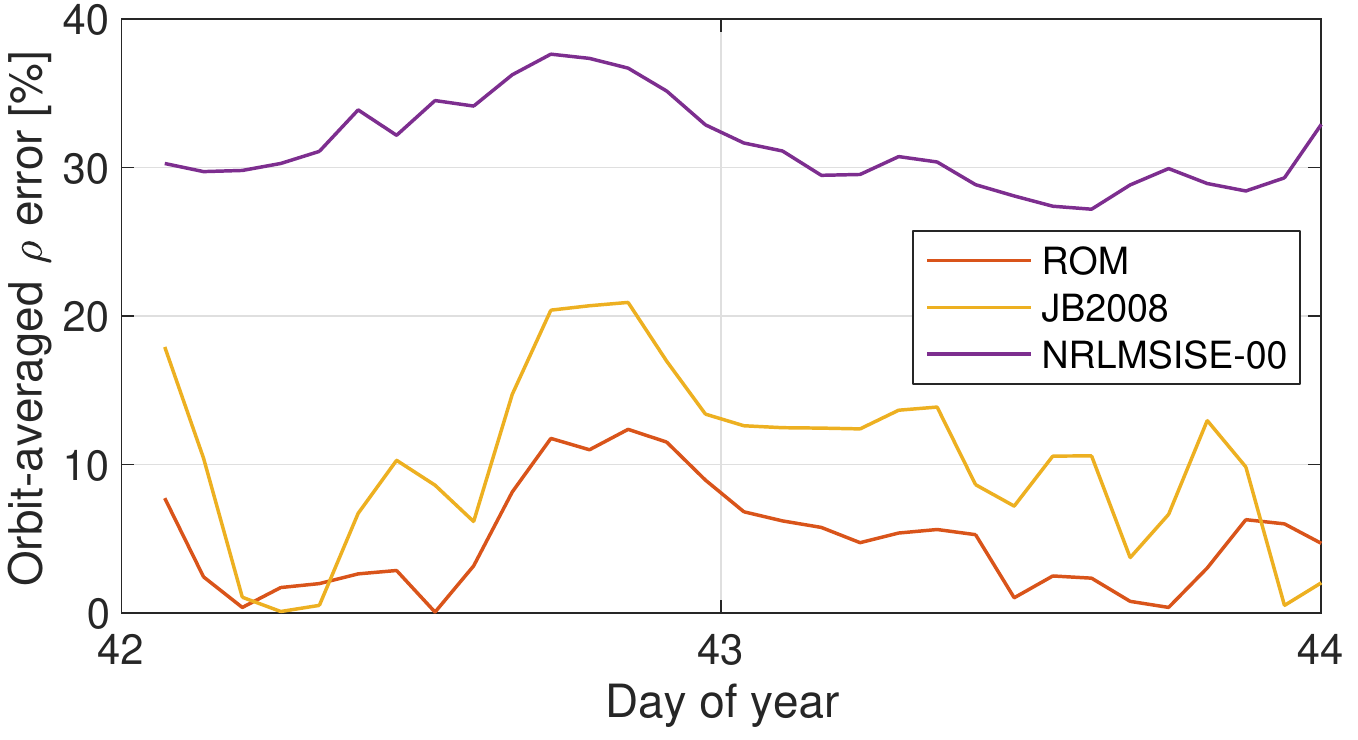} \label{fig:champDensityPred_avgErr}}
     
     \caption{Density along CHAMP orbit according to CHAMP accelerometer-derived data, ROM prediction, and JB2008 and NRLMSISE-00 models from February 11 to 13, 2003.}
     \label{fig:champDensityPred}
\end{figure}

To demonstrate the density and uncertainty prediction, we compared the predicted density and uncertainty along CHAMP's orbit with CHAMP accelerometer-derived densities. Figure~\ref{fig:champDensityPred} shows the 2-day predicted density along CHAMP's orbit (at about 415 km altitude) together with NRLMSISE-00, JB2008 and accelerometer-derived densities. 
The densities predicted using the ROM density model are very accurate with a maximum orbit-averaged density error of 12\%, see Figure~\ref{fig:champDensityPred_avgErr}. On the other hand, Figures \ref{fig:champDensityPred} and \ref{fig:champDensityPred_avg} show that the density according to the NRLMSISE-00 model is significantly biased in this time window. The ROM estimated density and 3-$\sigma$ uncertainty are shown in Figure~\ref{fig:champDensityPred_cov}. In the 2-day window, 56\% of the accelerometer-derived densities are within the 3-$\sigma$ bounds. This means that the true error in local density is larger than estimated. The uncertainty in the local density is underestimated because the short-term fluctuations in density cannot be detected from TLE data, which is generated from multi-day observation data. 

\subsection{Gaussianity}
\begin{figure}
     \centering
     \includegraphics[width=0.6\textwidth,trim={0cm 0cm 0cm 0cm},clip]{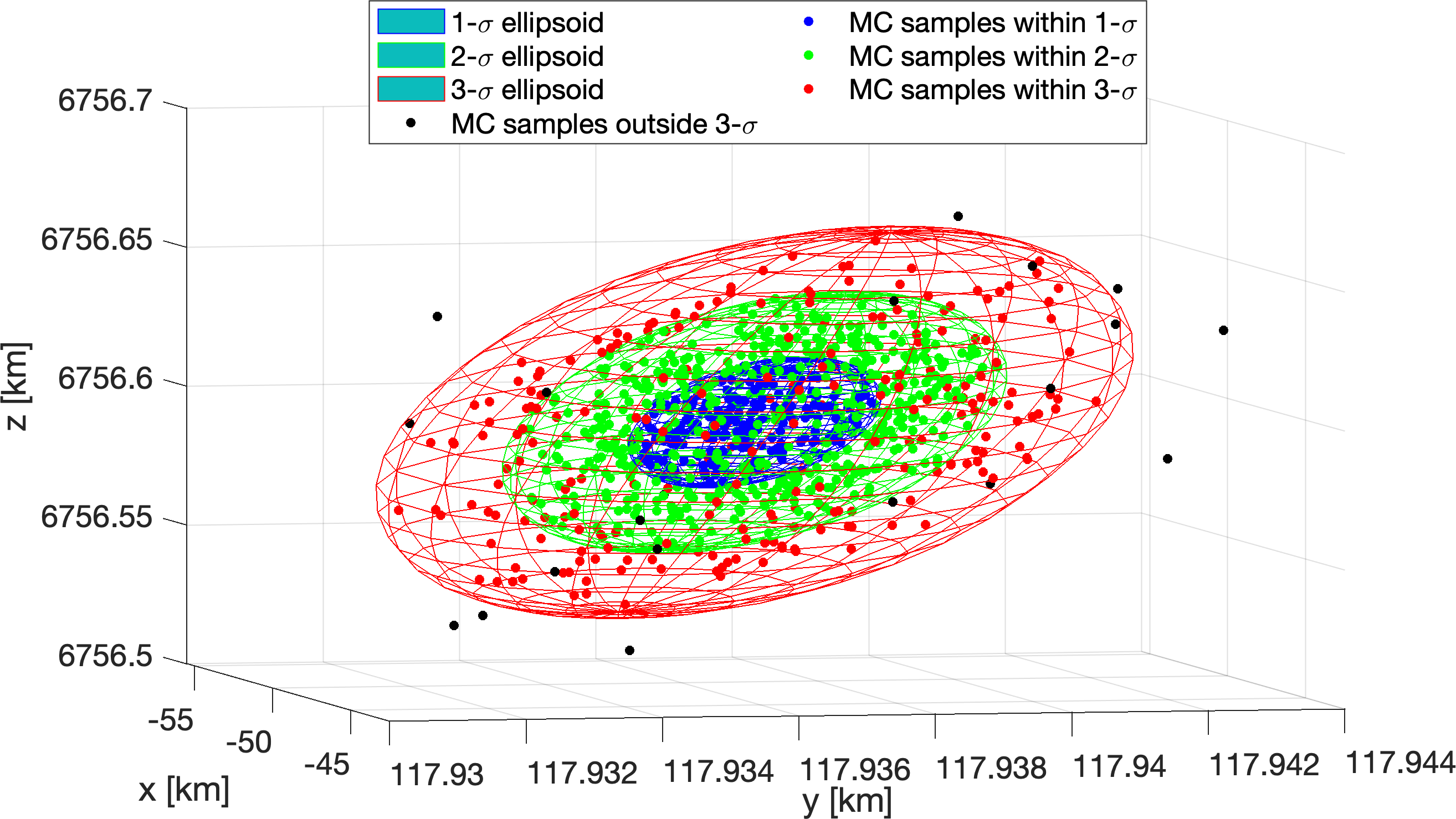}
     \caption{Monte Carlo samples versus 1-$\sigma$, 2-$\sigma$ and 3-$\sigma$ ellipsoids according to position covariance matrix.}
     \label{fig:gaussianity}
\end{figure}

For accurate estimation of the $P_c$ using Alfano's method, the position covariance needs to be Gaussian. We performed a Monte Carlo analysis to assess the Gaussianity of one of the position PDFs in test case S0. For this, we sampled 1000 initial conditions from the initial Gaussian distributions for the state, BC and reduced-order density, propagated them for two days and compared the final position with the position uncertainty according the propagated covariance matrix.
Figure~\ref{fig:gaussianity} shows the 1, 2 and 3-$\sigma$ ellipsoids computed using propagated position covariance matrix and 1000 Monte Carlo samples. The 3-$\sigma$ ellipsoid in this figure corresponds to the $P_1$ ellipsoid in Figures~\ref{fig:covEllipsWithDens} and \ref{fig:covEllipsWithCorr}. The Gaussian PDF provided by covariance matrix seems a good approximation of the true PDF. This is confirmed by Table~\ref{tab:gaussianity} that shows the fraction of Monte Carlo samples within the $\sigma$ bounds and the expected fraction for a three-dimensional Gaussian distribution. The expected and true fraction differ by maximum 1.3\%, so the Gaussian assumption for the position PDF is valid. In future work, other multivariate normality tests can be employed to assess the Gaussianity of pdfs \cite{MARDIA1980tests}.

\subsection{Probability of collision including density uncertainty}

\begin{table}
\centering
\begin{tabular}{ccc}
\hline
Range & Expected fraction of & Fraction of MC  \\
 & samples inside range & samples inside range \\
 \hline
1-$\sigma$ & 0.1987 & 0.200 \\
2-$\sigma$ & 0.7385 & 0.748 \\
3-$\sigma$ & 0.9707 & 0.979 \\
\hline
\end{tabular}
\caption{Fraction of Monte Carlo (MC) samples inside 1-$\sigma$, 2-$\sigma$ and 3-$\sigma$ range according to unscented covariance propagation. MC using 1000 samples, sampled from normal distributions in MEE, BC, and ROM.}
\label{tab:gaussianity}
\end{table}

\begin{figure}
     \centering
     \subfigure[][No density uncertainty]{\includegraphics[width=0.47\textwidth,trim={0 0 0 0},clip]{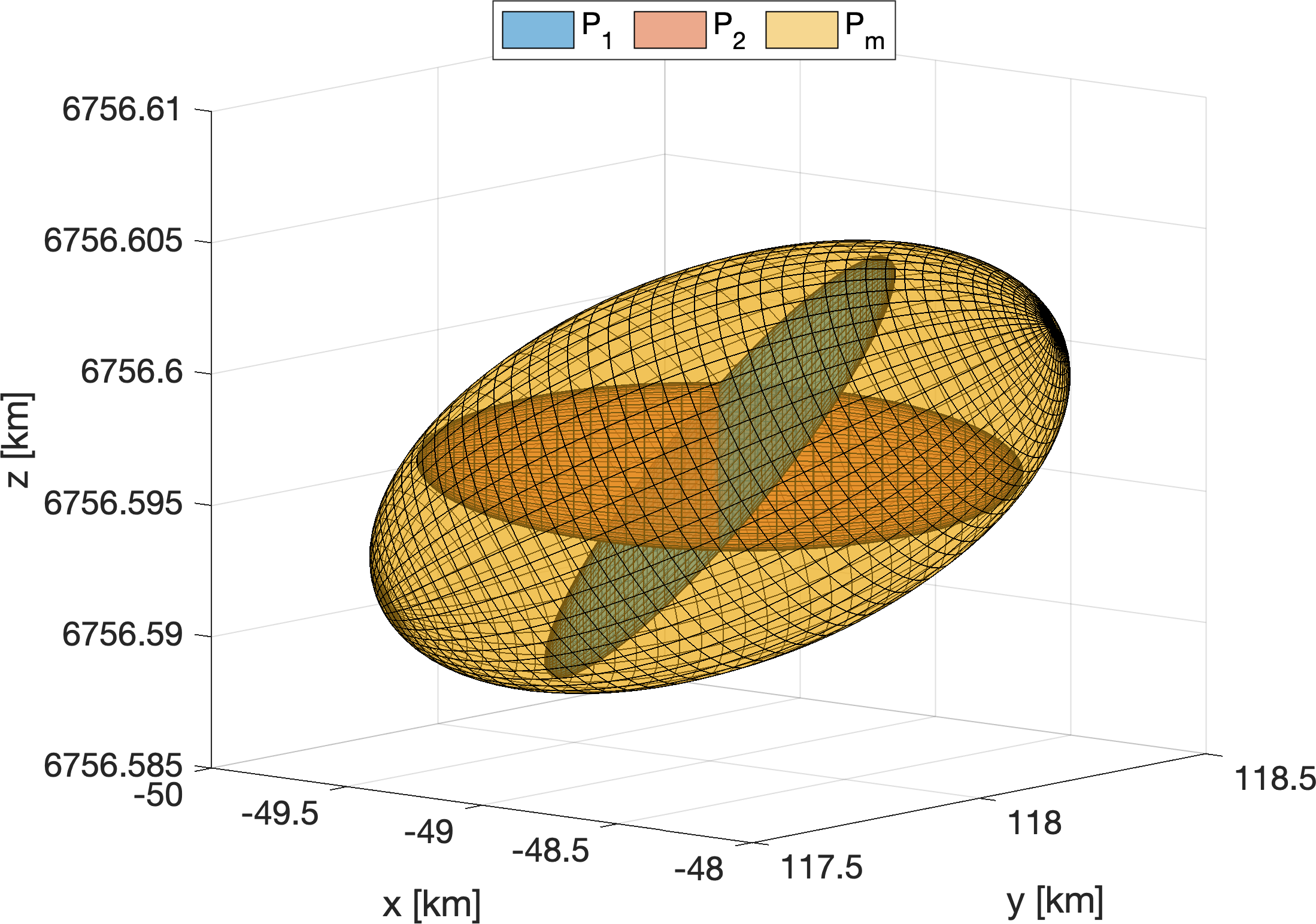} \label{fig:covEllipsNoDens}}
     
     \subfigure[][Including density uncertainty]{\includegraphics[width=0.47\textwidth,trim={0 0 0 0},clip]{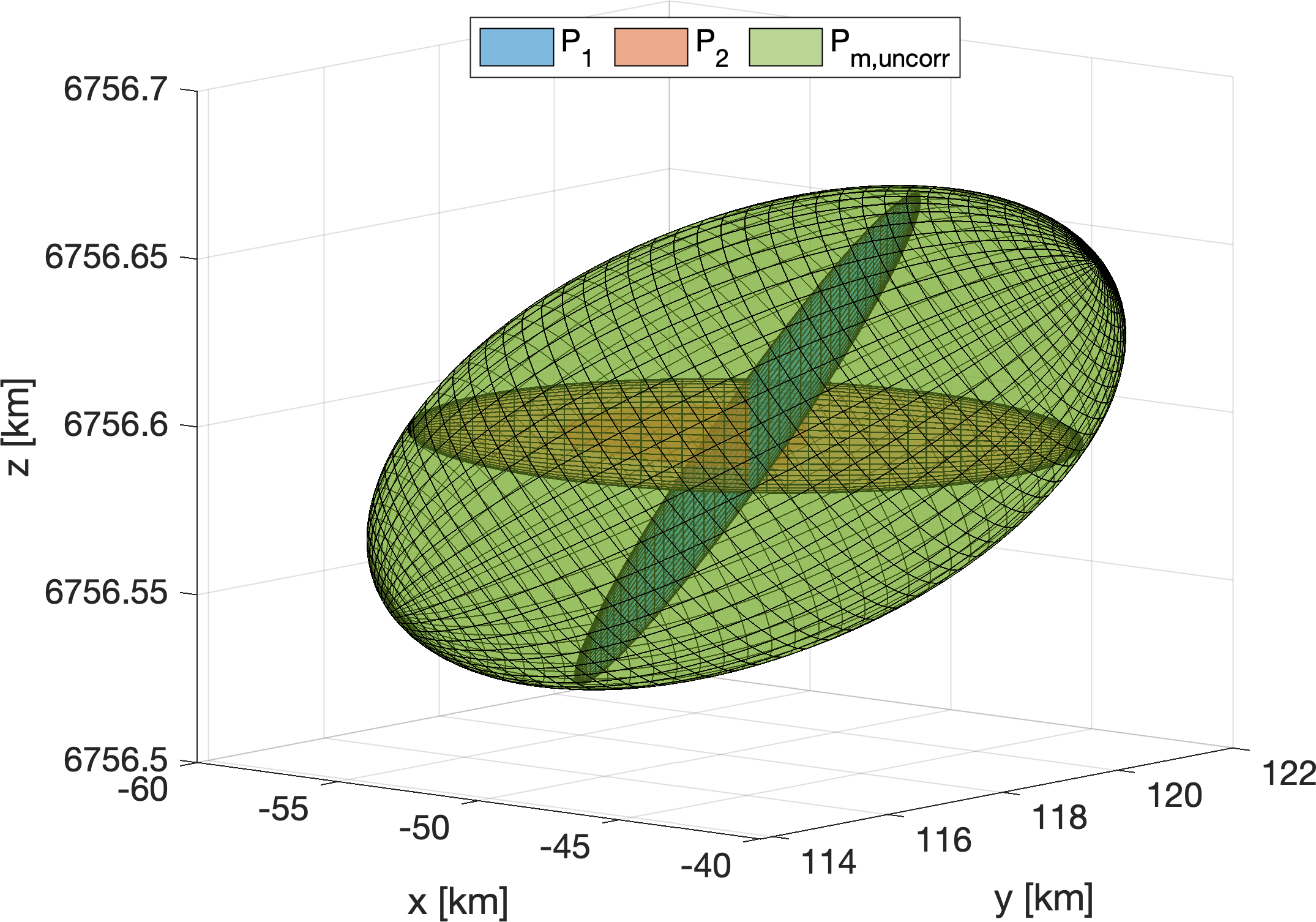} \label{fig:covEllipsWithDens}}
     \subfigure[][Including density uncertainty and cross-correlation]{\includegraphics[width=0.47\textwidth,trim={0 0 0 0},clip]{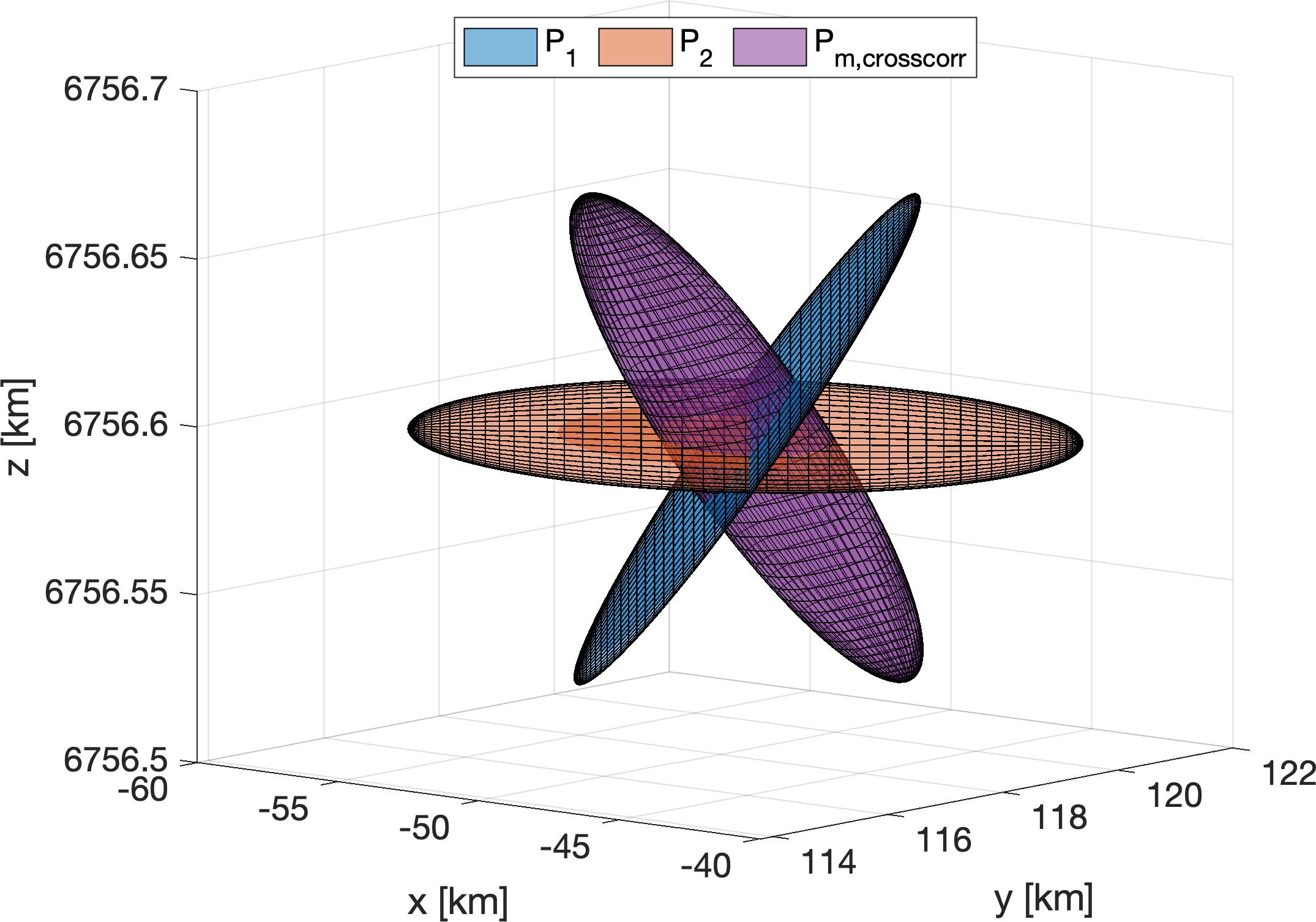} \label{fig:covEllipsWithCorr}}
     
     \caption{3-$\sigma$ ellipsoids for position covariances $\mathbf{P}_1$ and $\mathbf{P}_2$ and combined covariance $\mathbf{P}_
     m$ at TCA for test case S0: a) without density uncertainty; b) with density uncertainty that is assumed to be uncorrelated; c) with density uncertainty and considering cross-correlation.}
     \label{fig:covEllips}
\end{figure}

Using the propagated covariance for the position, we computed the $P_c$ including density uncertainty. Figure~\ref{fig:covEllips} shows the 3-$\sigma$ ellipsoids depicting the position covariance for objects 1 and 2 and the combined covariance at TCA for test case S0. Figure~\ref{fig:covEllipsNoDens} shows the covariance without considering density uncertainty, whereas Figures~\ref{fig:covEllipsWithDens} and Figure~\ref{fig:covEllipsWithCorr} include density uncertainty. In addition, the combined covariance in Figure~\ref{fig:covEllipsWithCorr} accounts for cross-correlation, whereas the combined covariance in Figure~\ref{fig:covEllipsWithDens} does not. As expected, the covariance get inflated when uncertainties due to density are included (note the different scales of the axes in Figure~\ref{fig:covEllipsNoDens} and in Figures~\ref{fig:covEllipsWithDens} and \ref{fig:covEllipsWithCorr}). Furthermore, Figure~\ref{fig:covEllipsWithCorr} shows that accounting for the cross-correlation can have a large effect on the combined covariance (i.e. the covariance of the miss distance) by deflating it. 

Table~\ref{tab:Pctable} shows the $P_c$ computed for the different conjunction scenarios with and without considering uncertainties in density. Let us first consider column two and three that show the $P_c$ in case of ignoring and including the density uncertainty without considering cross-correlation. We can see that due to the inflated covariance when including density uncertainties, the $P_c$ including density errors, $P_{c}^{\rho}$, is smaller than the $P_c$ without density errors, $P_{c}^{w/o}$, for very close approaches (see cases S0 and D0) and $P_{c}^{\rho}$ is larger than $P_{c}^{w/o}$ for more distance approaches (see cases S1, S2, D1 and D2). This effect of density uncertainty on the $P_c$ is as expected, see e.g. \cite{Hejduk2018Effect}.

\begin{table}
\centering
\begin{tabular}{lclcccc}
\hline
Scenario & Test & DCA & \multicolumn{4}{c}{Probability of collision ($P_c$)} \\
\cline{4-7}
 & case & [km] & No density & \multicolumn{3}{c}{With density uncertainty} \\
\cline{5-7}
 &  &   & uncertainty & No correlation & Cross-correlation & MC - Mean \\
 &  &   & $P_{c}^{w/o}$ & $P_{c}^{\rho}$ & $P_{c}^{\rho,corr}$ & $P_{c}^{\rho,MC}$ \\
\hline
\multirow{3}{*}{Same BC} & S0 & 0.00031 & 4.0274E-03 & 9.3303E-05 & 1.7237E-03 &	2.3648E-03 \\
 & S1 & 0.92375 & 1.0286E-07 & 7.1163E-05 & 2.6821E-05 & 9.6136E-07 \\
 & S2 & 1.84755 & 2.3178E-22 & 3.1567E-05 & 6.6025E-11 & 2.1842E-17 \\
\hline
\multirow{3}{*}{Different BC} & D0 & 0.00015 & 5.3254E-03 & 1.6650E-04 & 2.0025E-04 & 2.2915E-04 \\
 & D1 & 0.92412 & 2.0371E-09 & 1.0200E-04 & 1.1094E-04 & 8.5594E-05 \\
 & D2 & 1.84792 & 0.0 & 2.3462E-05 & 1.8872E-05 & 1.9817E-05 \\
\hline
\end{tabular}
\caption{Probability of collision ($P_c$) for two BC scenarios and for different distances of close approach computed with and without considering density uncertainties and cross-correlation, and computed using Monte Carlo (MC) analysis.}
\label{tab:Pctable}
\end{table}

\subsection{Effect of cross-correlation}
The position covariances of the two objects are correlated because they fly through the same atmosphere. Therefore, we computed the $P_c$ that accounts for the effect of cross-correlation, $P_{c}^{\rho,corr}$, as described in Section~\ref{sec:crosscorr}. The resulting $P_c$ is shown in column four in Table~\ref{tab:Pctable}. $P_{c}^{\rho,corr}$ is larger than $P_{c}^{\rho}$ for very close approaches (see cases S0 and D0) and smaller than $P_{c}^{\rho}$ for more distance approaches. This is as expected because accounting for the cross-correlation deflates the covariance.

\begin{figure}
     \centering
     \includegraphics[width=0.85\textwidth,trim={0cm 0cm 0cm 0cm},clip]{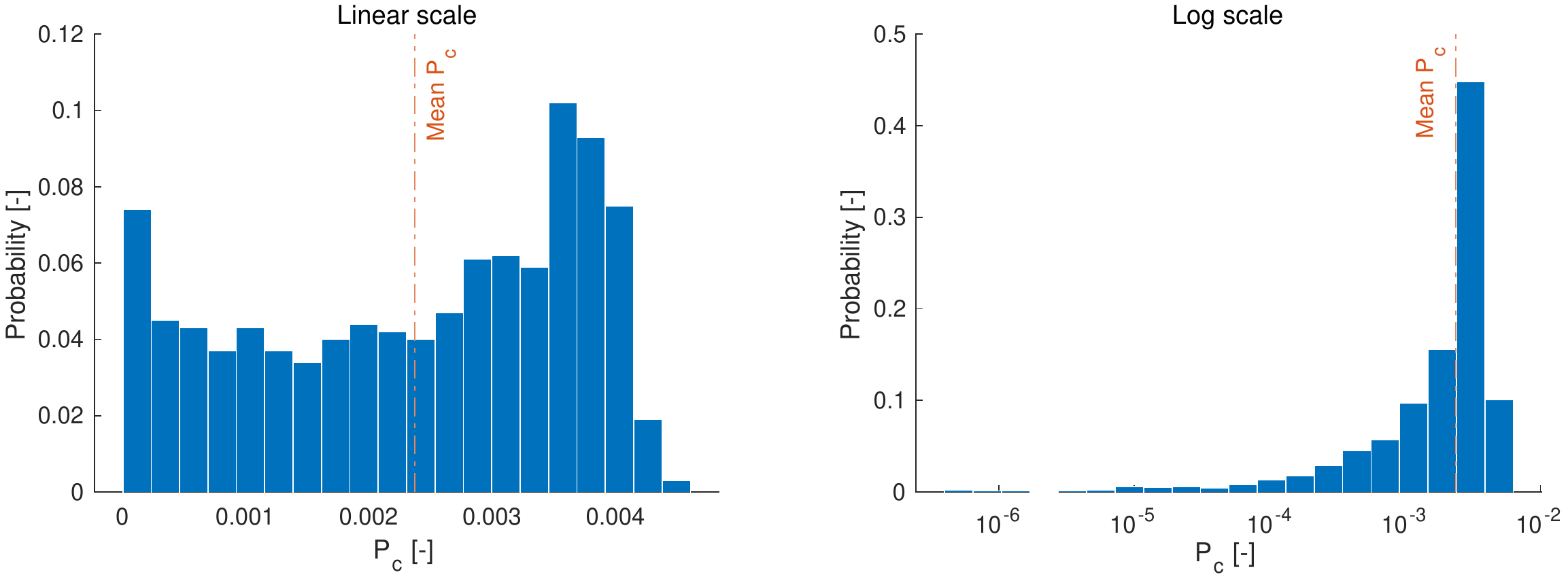}
     \caption{Probability distribution function of $P_c$ as a result of uncertainties in the atmospheric density for test case S0.}
     \label{fig:Pc_RomMC_PDF}
\end{figure}

To verify the calculation of the $P_c$ considering cross-correlation, we performed a Monte Carlo analysis to compute the expected $P_c$ considering atmospheric density uncertainty. One thousand random initial density states were sampled from the initial density PDF and for each sample the $P_c$ considering only initial state errors was computed. Figure~\ref{fig:Pc_RomMC_PDF} shows the probability distribution of the $P_c$ for test case S0. The mean $P_c$ was found to be 0.0023648. For the same test case, $P_{c}^{\rho,corr}$ computed considering density uncertainty and cross-correlation is 0.0017237, whereas $P_{c}^{\rho}$ without considering cross-correlation due to density uncertainty is 9.330E-05 and $P_{c}^{w/o}$ without considering uncertainty in the density is 0.004027. Clearly, correcting for cross-correlation improves the $P_c$ estimate. In particular, for such a high $P_c$ case, it is important to take cross-correlation into account, because the $P_c$ without considering cross-correlation is severely underestimating the realistic $P_c$. On the other hand, for the larger miss distances (case S2 and D2), accounting for cross-correlation results in a lower $P_c$ because of deflation of the combined covariance. These findings are in agreement with Casali et al. \cite{casali2018effect}.

For the conjunction scenarios where the two objects have a different BC, the effect of cross-correlation on the $P_c$ is much smaller. This is because the effect of drag on object 2 is smaller due to a smaller BC and therefore the effect of density errors on the $P_c$ is smaller and the cross-correlation is smaller. Still, for the high $P_c$ case D0, the cross-correlation correction improves the collision probability estimate; $P_{c}^{\rho,corr}$ is closer to the estimated true $P_c$ (given by $P_{c}^{\rho,MC}$) than $P_{c}^{\rho}$. In all cases except case D1, accounting for cross-correlation results in a better estimate of the $P_c$ (i.e. closer to $P_{c}^{\rho,MC}$). It can be noted that in case D1 the effect of correlation is very small, $P_{c}^{\rho}$ is close to $P_{c}^{\rho,MC}$.
These results show that 1) one must compensate for the cross-correlation due to density errors and 2) our approach to account for the effect of cross-correlation provides improved $P_c$ estimates.

\section{Conclusions}
In this paper we have demonstrated how atmospheric density uncertainties can be estimated and included for conjunction assessment. We used the recently developed dynamic reduced-order density model to estimate the density using two-line element data and quantified the uncertainty in the estimates using a Kalman filter. The benefit of this approach is that 1) we can quantify the uncertainty in the density, 2) we can propagate the uncertainty forward and 3) the uncertainty is location (i.e. latitude, longitude, altitude) and time (i.e. solar activity) dependent.

The estimated uncertainties in the density were included for collision probability calculation by propagating the state and density uncertainties simultaneously to obtain the position covariance at TCA. In addition, because the position covariances of two objects flying through the same atmosphere are correlated, we accounted for the effect of cross-correlation due to density errors on the $P_c$. 

The density uncertainty was shown to have a significant effect on the $P_c$. Including density uncertainty inflates the position covariances and results in a lower $P_c$ for very close approaches and in a higher $P_c$ for more distance approaches compared to ignoring errors due to density. Moreover, the results showed that it is important to consider the effect of cross-correlation on the $P_c$, especially when the effect of density errors on the orbit is similar for both objects. Ignoring the cross-correlation for very close approaches can result in severe underestimation of the collision probability.

The presented approach provides the distinctive capability to quantify the uncertainty in atmospheric density and to include this uncertainty for conjunction assessment while taking into account the dependence of the density model errors on location and time.

Finally, in this work we focused on the effect of density model errors, which is dominant for short-term orbit prediction. In future work, the effect of space weather prediction errors on the $P_c$ can be included as well.

\section*{Acknowledgments}
The authors wish to acknowledge support of this work by the Air Force's Office of Scientific Research under Contract Number FA9550-18-1-0149 issued by Erik Blasch. The NRLMSISE-00 and JB2008 models used in this work can be found on \url{https://www.brodo.de/space/nrlmsise} and \url{http://sol.spacenvironment.net/jb2008/code.html}, respectively. The space weather proxies were obtained from \url{http://celestrak.com/SpaceData/} and \url{http://sol.spacenvironment.net/jb2008/indices.html}. The CHAMP and GRACE densities used in this work were derived by \citet{sutton2008effects} and can be found, e.g., on \url{http://tinyurl.com/densitysets}.

\bibliography{references}

\begin{thebibliography}{31}
\newcommand{\enquote}[1]{``#1''}
\providecommand{\natexlab}[1]{#1}
\providecommand{\url}[1]{\texttt{#1}}
\providecommand{\urlprefix}{URL }
\expandafter\ifx\csname urlstyle\endcsname\relax
  \providecommand{\doi}[1]{doi:\discretionary{}{}{}#1}\else
  \providecommand{\doi}{doi:\discretionary{}{}{}\begingroup
  \urlstyle{rm}\Url}\fi

\bibitem[{Vallado and Finkleman(2014)}]{vallado2014critical}
Vallado, D.~A., and Finkleman, D., \enquote{A critical assessment of satellite
  drag and atmospheric density modeling,} \emph{Acta Astronautica}, Vol.~95,
  2014, pp. 141--165.

\bibitem[{He et~al.(2018)He, Yang, Carter, Kerr, Wu, Deleflie, Cai, Zhang,
  Sagni\`{e}res, and Norman}]{he2018review}
He, C., Yang, Y., Carter, B., Kerr, E., Wu, S., Deleflie, F., Cai, H., Zhang,
  K., Sagni\`{e}res, L., and Norman, R., \enquote{Review and comparison of
  empirical thermospheric mass density models,} \emph{Progress in Aerospace
  Sciences}, Vol. 103, 2018, pp. 31 -- 51.
\newblock \doi{https://doi.org/10.1016/j.paerosci.2018.10.003}.

\bibitem[{Poore et~al.(2016)Poore, Aristoff, Horwood, Armellin, Cerven, Cheng,
  Cox, Erwin, Frisbee, Hejduk et~al.}]{poore2016covariance}
Poore, A.~B., Aristoff, J.~M., Horwood, J.~T., Armellin, R., Cerven, W.~T.,
  Cheng, Y., Cox, C.~M., Erwin, R.~S., Frisbee, J.~H., Hejduk, M.~D., et~al.,
  \enquote{Covariance and uncertainty realism in space surveillance and
  tracking,} Tech. rep., Numerica Corporation Fort Collins United States, 2016.

\bibitem[{Anderson et~al.(2009)Anderson, Born, and
  Forbes}]{anderson2009sensitivity}
Anderson, R.~L., Born, G.~H., and Forbes, J.~M., \enquote{Sensitivity of orbit
  predictions to density variability,} \emph{Journal of Spacecraft and
  Rockets}, Vol.~46, No.~6, 2009, pp. 1214--1230.

\bibitem[{Leonard et~al.(2012)Leonard, Forbes, and Born}]{leonard2012impact}
Leonard, J., Forbes, J., and Born, G., \enquote{Impact of tidal density
  variability on orbital and reentry predictions,} \emph{Space Weather},
  Vol.~10, No.~12, 2012, pp. 1--12.

\bibitem[{Emmert et~al.(2014)Emmert, Byers, Warren, and
  Segerman}]{emmert2014propagation}
Emmert, J., Byers, J., Warren, H., and Segerman, A., \enquote{Propagation of
  forecast errors from the Sun to LEO trajectories: How does drag uncertainty
  affect conjunction frequency?} Tech. rep., NAVAL RESEARCH LAB WASHINGTON DC,
  2014.

\bibitem[{Emmert et~al.(2017)Emmert, Warren, Segerman, Byers, and
  Picone}]{emmert2017propagation}
Emmert, J., Warren, H., Segerman, A., Byers, J., and Picone, J.,
  \enquote{Propagation of atmospheric density errors to satellite orbits,}
  \emph{Advances in Space Research}, Vol.~59, No.~1, 2017, pp. 147--165.

\bibitem[{Sagnieres and Sharf(2017)}]{sagnieres2017uncertainty}
Sagnieres, L., and Sharf, I., \enquote{Uncertainty characterization of
  atmospheric density models for orbit prediction of space debris,} \emph{7th
  European Conference on Space Debris}, 2017.

\bibitem[{Schiemenz et~al.(2019{\natexlab{a}})Schiemenz, Utzmann, and
  Kayal}]{SCHIEMENZ2019grid}
Schiemenz, F., Utzmann, J., and Kayal, H., \enquote{Propagation of grid-scale
  density model uncertainty to orbital uncertainties,} \emph{Advances in Space
  Research}, 2019{\natexlab{a}}.
\newblock \doi{https://doi.org/10.1016/j.asr.2019.10.013},
  \urlprefix\url{http://www.sciencedirect.com/science/article/pii/S0273117719307471}.

\bibitem[{Schiemenz et~al.(2019{\natexlab{b}})Schiemenz, Utzmann, and
  Kayal}]{SCHIEMENZ2019Least}
Schiemenz, F., Utzmann, J., and Kayal, H., \enquote{Least squares orbit
  estimation including atmospheric density uncertainty consideration,}
  \emph{Advances in Space Research}, Vol.~63, No.~12, 2019{\natexlab{b}}, pp.
  3916 -- 3935.
\newblock \doi{https://doi.org/10.1016/j.asr.2019.02.039},
  \urlprefix\url{http://www.sciencedirect.com/science/article/pii/S0273117719301656}.

\bibitem[{Schiemenz et~al.(2019{\natexlab{c}})Schiemenz, Utzmann, and
  Kayal}]{SCHIEMENZ2019Prop}
Schiemenz, F., Utzmann, J., and Kayal, H., \enquote{Propagating EUV solar flux
  uncertainty to atmospheric density uncertainty,} \emph{Advances in Space
  Research}, Vol.~63, No.~12, 2019{\natexlab{c}}, pp. 3936 -- 3952.
\newblock \doi{https://doi.org/10.1016/j.asr.2019.02.040},
  \urlprefix\url{http://www.sciencedirect.com/science/article/pii/S027311771930167X}.

\bibitem[{Cefola et~al.(2004)Cefola, Proulx, Nazarenko, and
  Yurasov}]{cefola2004atmospheric}
Cefola, P.~J., Proulx, R.~J., Nazarenko, A.~I., and Yurasov, V.~S.,
  \enquote{Atmospheric density correction using two line element sets as the
  observation data,} \emph{Advances in the Astronautical Sciences}, Vol. 116,
  2004, pp. 1953--1978.

\bibitem[{Yurasov et~al.(2005)Yurasov, Nazarenko, Cefola, and
  Alfriend}]{yurasov2005density}
Yurasov, V.~S., Nazarenko, A.~I., Cefola, P.~J., and Alfriend, K.~T.,
  \enquote{Density corrections for the {NRLMSIS}-00 atmosphere model,}
  \emph{{Proceedings of the AAS/AIAA Space Flight Mechanics Conference, January
  23-27, Copper Mountain, CO}}, 2005, pp. 1079--1107.

\bibitem[{Storz et~al.(2005)Storz, Bowman, Branson, Casali, and
  Tobiska}]{storz2005high}
Storz, M.~F., Bowman, B.~R., Branson, M. J.~I., Casali, S.~J., and Tobiska,
  W.~K., \enquote{High accuracy satellite drag model (HASDM),} \emph{Advances
  in Space Research}, Vol.~36, No.~12, 2005, pp. 2497--2505.

\bibitem[{Doornbos et~al.(2008)Doornbos, Klinkrad, and
  Visser}]{doornbos2008use}
Doornbos, E., Klinkrad, H., and Visser, P., \enquote{Use of two-line element
  data for thermosphere neutral density model calibration,} \emph{Advances in
  Space Research}, Vol.~41, No.~7, 2008, pp. 1115--1122.

\bibitem[{Chen et~al.(2019)Chen, Du, and Sang}]{chen2019improved}
Chen, J., Du, J., and Sang, J., \enquote{Improved orbit prediction of {LEO}
  objects with calibrated atmospheric mass density model,} \emph{Journal of
  Spatial Science}, Vol.~64, No.~1, 2019, pp. 97--110.

\bibitem[{Bussy-Virat et~al.(2018)Bussy-Virat, Ridley, and
  Getchius}]{bussy2018effects}
Bussy-Virat, C.~D., Ridley, A.~J., and Getchius, J.~W., \enquote{Effects of
  uncertainties in the atmospheric density on the probability of collision
  between space objects,} \emph{Space Weather}, Vol.~16, No.~5, 2018, pp.
  519--537.

\bibitem[{Hejduk and Snow(2018)}]{Hejduk2018Effect}
Hejduk, M.~D., and Snow, D.~E., \enquote{The Effect of Neutral Density
  Estimation Errors on Satellite Conjunction Serious Event Rates,} \emph{Space
  Weather}, Vol.~16, No.~7, 2018, pp. 849--869.
\newblock \doi{10.1029/2017SW001720},
  \urlprefix\url{https://agupubs.onlinelibrary.wiley.com/doi/abs/10.1029/2017SW001720}.

\bibitem[{Coppola et~al.(2004)Coppola, Woodburn, and
  Hujsak}]{coppola2004effects}
Coppola, V.~T., Woodburn, J., and Hujsak, R., \enquote{Effects of Cross
  Correlated Covariance on Space-Craft Collision Probability,} \emph{AAS/AIAA
  Spaceflight Mechanics Meeting}, 2004, pp. 04--181.

\bibitem[{Casali et~al.(2018)Casali, Hall, Snow, Hejduk, Johnson, Skrehart, and
  Baars}]{casali2018effect}
Casali, S., Hall, D., Snow, D., Hejduk, M., Johnson, L., Skrehart, B., and
  Baars, L., \enquote{Effect of Cross-Correlation of Orbital Error on
  Probability of Collision Determination,} \emph{AAS/AIAA Astrodynamics
  Specialist Conference Paper}, 2018, pp. 18--272.

\bibitem[{Mehta et~al.(2018)Mehta, Linares, and Sutton}]{mehta2018quasi}
Mehta, P.~M., Linares, R., and Sutton, E.~K., \enquote{A Quasi-Physical Dynamic
  Reduced Order Model for Thermospheric Mass Density via Hermitian
  Space-Dynamic Mode Decomposition,} \emph{Space Weather}, Vol.~16, No.~5,
  2018, pp. 569--588.
\newblock \doi{https://doi.org/10.1029/2018SW001840}.

\bibitem[{Mehta and Linares(2018{\natexlab{a}})}]{mehta2018new}
Mehta, P.~M., and Linares, R., \enquote{A New Transformative Framework for Data
  Assimilation and Calibration of Physical Ionosphere-Thermosphere Models,}
  \emph{Space Weather}, Vol.~16, No.~8, 2018{\natexlab{a}}, pp. 1086--1100.
\newblock \doi{https://doi.org/10.1029/2018SW001875}.

\bibitem[{Mehta and Linares(2018{\natexlab{b}})}]{mehta2018data}
Mehta, P.~M., and Linares, R., \enquote{Data-driven framework for real-time
  thermospheric density estimation,} \emph{{Proceedings of the AAS/AIAA
  Astrodynamics Specialist Conference, August 19-23, Snowbird, UT}},
  2018{\natexlab{b}}.

\bibitem[{Gondelach and Linares(2019)}]{gondelach2019real}
Gondelach, D.~J., and Linares, R., \enquote{Real-Time Thermospheric Density
  Estimation Via Two-Line-Element Data Assimilation,} \emph{arXiv preprint
  arXiv:1910.00695}, 2019.

\bibitem[{Mehta and Linares(2017)}]{mehta2017methodology}
Mehta, P.~M., and Linares, R., \enquote{A methodology for reduced order
  modeling and calibration of the upper atmosphere,} \emph{Space Weather},
  Vol.~15, No.~10, 2017, pp. 1270--1287.
\newblock \doi{https://doi.org/10.1002/2017SW001642}.

\bibitem[{Wan and Van Der~Merwe(2001)}]{wan2001unscented}
Wan, E.~A., and Van Der~Merwe, R., \enquote{The unscented {Kalman} filter,}
  \emph{Kalman filtering and neural networks}, John Wiley and Sons, 2001,
  Chap.~7, pp. 221--280.

\bibitem[{Alfano(2005)}]{alfano2005numerical}
Alfano, S., \enquote{A numerical implementation of spherical object collision
  probability,} \emph{Journal of Astronautical Sciences}, Vol.~53, No.~1, 2005,
  pp. 103--109.

\bibitem[{Hemenway et~al.(2014)Hemenway, Welser, and
  Baiocchi}]{Hemenway2014Achieving}
Hemenway, B., Welser, W., and Baiocchi, D., \enquote{Achieving Higher-Fidelity
  Conjunction Analyses Using Cryptography to Improve Information Sharing,}
  Tech. rep., Santa Monica, CA: RAND Corporation, 2014.

\bibitem[{Julier and Uhlmann(1997)}]{julier1997new}
Julier, S.~J., and Uhlmann, J.~K., \enquote{New extension of the {Kalman}
  filter to nonlinear systems,} \emph{Signal processing, sensor fusion, and
  target recognition {VI}}, Vol. 3068, International Society for Optics and
  Photonics, 1997, pp. 182--193.
\newblock \doi{10.1117/12.280797}.

\bibitem[{Mardia(1980)}]{MARDIA1980tests}
Mardia, K., \enquote{Tests of unvariate and multivariate normality,}
  \emph{Analysis of Variance}, Handbook of Statistics, Vol.~1, Elsevier, 1980,
  pp. 279 -- 320.
\newblock \doi{https://doi.org/10.1016/S0169-7161(80)01011-5},
  \urlprefix\url{http://www.sciencedirect.com/science/article/pii/S0169716180010115}.

\bibitem[{Sutton(2008)}]{sutton2008effects}
Sutton, E.~K., \enquote{Effects of solar disturbances on the thermosphere
  densities and winds from {CHAMP} and {GRACE} satellite accelerometer data,}
  Ph.D. thesis, University of Colorado at Boulder, October 2008.

\end{thebibliography}

\end{document}